\documentclass{ifacconf}
\usepackage{algorithm}
\usepackage[noend]{algpseudocode}
\usepackage{enumerate}
\usepackage{graphicx}      % include this line if your document contains figures
\usepackage{natbib}        % required for bibliography
\usepackage{xcolor}
\usepackage{mathtools}  
\begin{document}
\begin{frontmatter}

\title{An Online Evolving Framework for Advancing Reinforcement-Learning based Automated Vehicle Control\thanksref{footnoteinfo}} 
\thanks[footnoteinfo]{This work was supported by the Ford Motor Company; © 2020 the authors. This work has been accepted to IFAC for publication under a Creative Commons Licence CC-BY-NC-ND}

\author[First]{Teawon Han} 
\author[Second]{Subramanya Nageshrao}
\author[Third]{Dimitar P. Filev} 
\author[First]{{\"U}mit {\"O}zg{\"u}ner}

\address[First]{Department of Electrical and Computer Engineering,
  The Ohio State University, Columbus, OH 43210, USA (e-mail: han.394@osu.edu, ozguner.1@osu.edu)}
\address[Second]{Ford Greenfield Labs, Palo Alto, CA 94304, USA (e-mail: snageshr@ford.com)}
\address[Third]{Ford Motor Company, Dearborn, MI 48121, USA (e-mail: dfilev@ford.com)}

\begin{abstract}                % Abstract of not more than 250 words.
In this paper, an online evolving framework is proposed to detect and revise a controller's imperfect decision-making in advance. The framework consists of three modules: the evolving Finite State Machine (e-FSM), action-reviser, and controller modules. The e-FSM module evolves a stochastic model (e.g., Discrete-Time Markov Chain) from scratch by determining new states and identifying transition probabilities repeatedly. With the latest stochastic model and given criteria, the action-reviser module checks validity of the controller's chosen action by predicting future states. Then, if the chosen action is not appropriate, another action is inspected and selected. In order to show the advantage of the proposed framework, the Deep Deterministic Policy Gradient (DDPG) w/ and w/o the online evolving framework are applied to control an ego-vehicle in the car-following scenario where control criteria are set by speed and safety. Experimental results show that inappropriate actions chosen by the DDPG controller are detected and revised appropriately through our proposed framework, resulting in no control failures after a few iterations.
\end{abstract}

\end{frontmatter}
%===============================================================================

%#####################################################################################
\section{Introduction}
\vspace{-0.2cm}
%#####################################################################################
For many decades, various methodologies have been proposed for safe or efficient automated vehicle (AV) control under different situations. Rule-based, supervised-learning, and reinforcement-learning are widely studied and applied to control AVs.

As a rule-based approach, the Finite State Machines (FSMs) which consist of the finite number of predefined states and transition-conditions has been implemented for decision-making under known situations. In the FSMs, either actions or control-strategies are described in each state, and the transitions occur when pre-determined conditions are met. So, the FSMs based controller returns the appropriate action based on the system's conditions. \cite{kurt2008hybrid} and \cite{redmill2008ohio} implement the Hierarchical FSMs as a high-level decision-maker within the Hybrid State System (HSS) to control automated vehicles (AVs) under the given several scenarios in the 2007 DARPA Urban Challenge. Although the AV (called OSU-ACT) is controlled via appropriate decision-making under complex situations, the capability of the rule-based controller is fully subordinate to the initially defined rules and conditions. For example, if unexpected situations are encountered, the controller can neither recognize the situations nor make the right decision due to no existing proper rules.

The Supervised Learning (SL) is focused on seeking the best coefficients or weights of the predefined control models by using a given cost-function and ground truth data. The data consists of pairs of inputs (situations or conditions) and outputs (optimal actions), and the optimal coefficients or weights are obtained by exploring the configuration which minimizes differences between the ground truth and the model’s output. In \cite{gadepally2013framework, kuge2000driver}, the Hidden Markov Model (HMM) is implemented for obtaining a general driver model that enables AV controllers to predict other vehicles’ behaviors, which is helpful for safer driving control. The HMM-based driver model is designed by the fixed number of hidden states and observations, and the transition and emission probabilities are identified via the Viterbi algorithm. In addition, various types of Deep Neural Network (DNN) are implemented to create AV control or driver model for predicting future traffic situations or choosing optimal actions respectively. \cite{han2019driving} proposes to train DNN by using raw observations and estimated general driving characteristics to predict other vehicle’s lane-change behavior on the highway. Also, the end-to-end learning-based driving controller, which maps raw camera-images with optimal controls via the Convolutional Neural Network (CNN), is derived by \cite{bojarski2016end}. \cite{al2017deep} implements the GoogLeNet to obtain accurate affordance parameters which are essential to take the optimal driving actions. Even though the SL approach has been widely used to get optimal models, a large amount of labeled data and off-line training procedure are required.

The Reinforcement Learning (RL) approach has been suggested to get an optimal controller which can improve its performance by updating its policy iteratively based on experiences. Unlike the SL, the ground truth data and pre-training process are not necessary, but a reward-function needs to be pre-defined appropriately. While the RL based controller explores or exploits actions at different states, its policy is updated repeatedly in a way to get the highest value (sum of future discounted rewards) at every state. The policy is represented by matrices or DNN. Since the DNN can well approximate value-function, various types of multilayer perceptron (MLP) have been proposed to represent the RL's policy. \cite{mnih2015human} proposes Deep Q Network (DQN), and \cite{van2016deep} suggests Double Deep Q-Network (DDQN) which can prevent the over-optimistic value estimation problem observed in the DQN. Since DQN and DDQN can only handle the discrete type of actions, Deep Deterministic Policy Gradient (DDPG) is proposed by \cite{lillicrap2015continuous} for the continuous type of actions. \cite{nageshrao2019autonomous} proposed the DRL agent control architecture, which implements DDQN for safe and adaptive AV control.

Due to the model-free attribution, the RL with MLP controller can find the optimal actions under unexpected situations, unlike the rule-based and SL approaches. However, a well-designed reward-function is still required. In this paper, instead of paying efforts for designing a perfect model or reward-function by hand, we propose an evolving methodology (called evolving Finite State Machine), which can derive a stochastic model from scratch by observing variations of system-conditions. Also, the derived stochastic model is used within the online evolving framework to check and revise the action which is chosen by the controller.

%Even though it is only shown that how the RL with MLP controller makes better decisions via the framework, any type of controller can be applied within the framework.

This paper is organized as follows: Section \ref{section:online evolving framework} describes the details of three modules within an on-line evolving framework. Section \ref{section:Target Scenario and Controller} introduces the DDPG AV controller and the car-following scenario settings. Section \ref{section:DDPG_controller_eFSM} explains the on-line evolving framework settings for the DDPG controller. Section \ref{section:experimental results} analyzes car-following performances of the DDPG controller w/ and w/o the suggested framework. Lastly, our contributions and further discussion are summarized in section \ref{section:conclusion}.

\section{An On-line Evolving Framework}
\label{section:online evolving framework}
\begin{figure}[ht]
\begin{center}
\includegraphics[width=7.5cm]{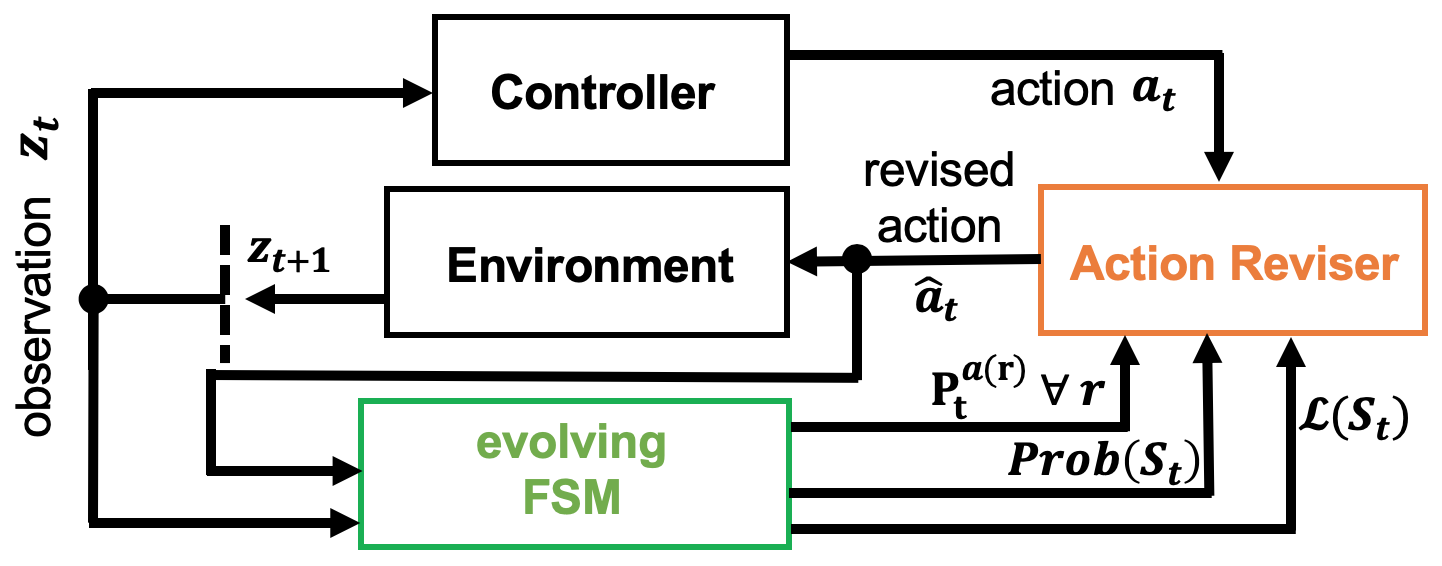}    % The printed column width is 8.4 cm.
\vspace{-0.2cm}
\caption{An on-line evolving framework for controllers} 
\label{fig:online_evolving_framework}
\end{center}
\end{figure}

\vspace{-0.2cm}
%#####################################################################################
An on-line evolving framework is proposed to inspect or revise an action selected by a controller. In order to inspect the action, it needs to predict future states depending on the action. In the framework shown in  Fig.~\ref{fig:online_evolving_framework}, there are a controller, evolving Finite State Machine (e-FSM), and action-reviser modules for choosing an optimal action, creating a stochastic model, and inspecting \& revising the controller's actions respectively.

A set of observations $z_t$ at time-step $t$ is fed into the controller and e-FSM modules. The control module makes a decision to choose the best action; the e-FSM module derives a stochastic model which consists of states and state-transition dynamics based on the applied actions; the action-reviser module checks validity of the chosen action $a_t$ and revises the action if necessary. Note that $a_t$, $Prob(S_t)$, $\mathbf{P}_{t}^{a(r)}\forall r$, and $\mathcal{L}(S_t)$ refer to an action chosen by the controller at time $t$, probability distributions over states (e.g., driving situations) which are determined by time $t$, identified state-transition probabilities with respect to actions, and a set of state attributions. In the following sub-sections, details of the e-FSM and the action-reviser modules are described.
% An e-FSM is introduced by \cite{han2019online}, which can determine and recognize system's states and identify state-transitions based on chosen actions. 
%#####################################################################################
\vspace{-0.2cm}
\subsection{evolving Finite State Machine(e-FSM)}
\label{subsec:e-FSM}
\begin{figure}[ht]
\begin{center}
\includegraphics[width=8cm]{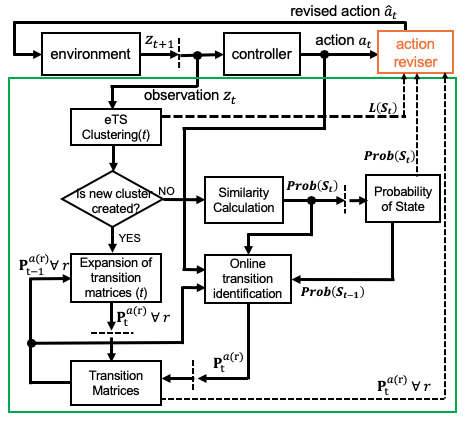}    % The printed column width is 8.4 cm.
\vspace{-0.2cm}
\caption{The framework of an evolving Finite State Machine} 
\label{fig:e_FSM_framework}
\end{center}
\end{figure}

\vspace{-0.2cm}
%#####################################################################################
The e-FSM, an on-line evolving method designed for supporting optimal decision-making, is a hybrid Markov model with states representing specific situations. The states are identified as clusters in the state space, and state dynamics is determined through a set of transition probability matrices associated with the inputs (actions). The framework of e-FSM is shown in Fig.~\ref{fig:e_FSM_framework}, and some of the key features (online state-determination, state flagging and recognition, and online transition-identification) are described in this paper; see \cite{han2019online} for e-FSM's further specific properties and descriptions. 
\vspace{-0.2cm}
\vspace{-0.1cm}
%#####################################################################################
\subsubsection{- Online State Determination }
\vspace{-0.3cm}
%#####################################################################################
In the e-FSM, states represent situations that a controller could be encountered. Due to the difficulty of pre-determining all possible situations, the e-FSM aims to determine a state whenever a new situation is observed over time. This feature enables the controller to have opportunities to find the best action by recognizing initially unexpected situations. 

To realize the online state determination, a set of observations $z_t=[o_t(1), ..., o_t(k)]^T$ at every time $t$ is clustered, and each cluster-center is considered as a state, representing individual situation. The evolving Takagi-Sugeno (eTS) online clustering method proposed by \cite{filev2010markov} is implemented so that infinite number of unknown states can be determined; the eTS can have infinite number of clusters. A state-set can be written by $\mathcal{S}_{t\geq1}=\{s_t(1), ..., s_t(n_t)\}$ where the states $s_t(i), 1\leq i\leq n_t$, are cluster-centers, and $n_t$ is the total number of the states determined by time $t$. This feature enables the structure of the stochastic model to evolve.
%#####################################################################################
\vspace{-0.25cm}
\subsubsection{- Flagging and Recognition of State}
\label{subsubsec:state labeling and recognition}
%#####################################################################################
The e-FSM determines states, in which each represents a distinguished situation. Instead of labeling all states, the states are flagged by either unfavorable states or not, where the unfavorable states are defined based on the given criteria. For example, if a given criterion of AV control is safety, the recognized state at the moment of collision is flagged as the unfavorable state. A set of the state attributions at time $t$, $\mathcal{L}(S_t)$, is transferred to the action-reviser as shown in Fig. \ref{fig:online_evolving_framework}, where $\mathcal{L}(s_t(i))\in\{none, unfavorable_j\}, 1\leq i \leq n_t$, $1\leq j\leq c_n$; $c_n$ is the total number of given criteria.

Given $z_t$, the current situation is represented by probability distributions over the states determined by time $t$, denoted as $Prob(S_t)=[Prob(s_t(1)), ..., Prob(s_t(n_t))]^T$, where each probability of state $Prob(s_t(\cdot))$ is obtained by calculating a similarity, $\lambda_t^i(z_t)$, between $z_t$ and $s_t(i), 1\leq i \leq n_t$ as defined in Eq.(\ref{eq:similarity}).
\begin{equation} \label{eq:similarity}
\begin{array}{ll}
&\lambda_t^i(z_t)=\frac{\eta_t^i(z_t)}{\sum_{j}\eta^{j}(z_{t})}  \\
& \textrm{where}\,\,\,  \eta_t^i(z_t)=exp\left(-\frac{(z_t-s_t^i)^T(z_t-s_t^i)}{var(s_t^i)}\right)
\end{array}
\end{equation}
%#####################################################################################
\vspace{-0.45cm}
\subsubsection{- Online State-Transitions Identification}
%#####################################################################################
The identification of state-transitions w.r.t. actions is important to predict future states precisely. For example, if an AV controller knows in advance what the future state would be based on the chosen action, it can get more chances to make a better decision. In the e-FSM, the state-transitions are represented by transition probability matrices (TPMs), of which each matrix is correlated to individual action in the action-set.

The action-set written by $\mathcal{A}_d=\{a(1), ..., a(q)\}$ is initially defined to have the fixed $q$ number of elements unlike the state-set $\mathcal{S}_t$ which is a variant set. The e-FSM uses discrete action-set but is compatible to controllers that have continuous action-set ($\mathcal{A}_c$) by encoding to discrete action-set ($\mathcal{A}_d$) with an arbitrary interval $\delta$. For example, the continuous action-set $\mathcal{A}_c=[-2.0\,\, 2.0]$ can be encoded with the interval $\delta=1.0$ to the discrete action-set such as $\mathcal{A}_d=\left\{a(1), a(2), a(3), a(4)\right\}$ where $a(1)=[-2.0\,\,-1.0)$, $a(2)=[-1.0\,\, 0.0)$, $a(3)=[0.0\,\, 1.0)$, $a(4)=[1.0\,\, 2.0]$.  

The state-transitions w.r.t. actions are represented by the $q$ number of TPMs. And, each TPM $\mathbf{P}_t^{a(r)}\forall r\in R^{{n_t}\times{n_t}}$ is correlated to each discrete action $a(r)\in \mathcal{A}_d$, where $1\leq r \leq q$ and $q$ is the total number of discrete actions in $\mathcal{A}_d$. The TPMs are defined by:
\begin{equation}\hspace{-0.3cm}
\begin{array}{ll}
    &\mathbf{P}_t^{a(r)}=\{P_t^{a(r)}(i,j)\}, 1\leq i,j \leq n_t\\
    &\textrm{where}\,\,\,P_t^{a(r)}(i,j)=Prob\left(s_{t+1}=s(j)|s_t=s(i), a_t=a(r)\right)
\end{array}
\end{equation}
When a new state is determined, the dimension of all TPMs is expanded by adding a column and a row, following the proposed method; see detail steps and an example given by \cite{han2019online}. Otherwise one of TPMs which is correlated to a chosen action is identified via Eq.(\ref{eq:transition_matrix_efsm}) and (\ref{eq:transition_matrix_efsm_where}) in a recursive way, designed by \cite{filev2010markov}. Note $\tau(t)$ and $\gamma(t)$ are probability distributions over states at time $t-1$ and $t$, $Prob(S_{t-1})$ and $Prob(S_t)$; $\varphi$ and $1_{n_t}$ are a learning rate and the $n_t$-dimensional vector of ones respectively. In the beginning, $F^{a(r)}(0)$ and $F_o^{a(r)}(0)$ are initialized by $\bar{\epsilon}$ which is a small non-negative constant. 
\begin{equation} \label{eq:transition_matrix_efsm}
        \mathbf{P}_t^{a(r)}=diag(F_0^{a(r)}(t))^{-1}F^{a(r)}(t)\\
\end{equation}
\begin{equation} \label{eq:transition_matrix_efsm_where}
% \hspace{-0.3cm}
\begin{array}{ll}
        &F^{a(r)}(t)=F^{a(r)}(t-1)+\varphi(\tau(t)\gamma(t)^T-F^{a(r)}(t-1))\\
        &F_o^{a(r)}(t)=F_o^{a(r)}(t-1)+\varphi(\tau(t)\gamma(t)^T1_{n_t}-F_o^{a(r)}(t-1))
\end{array}
\end{equation}
By using identified TPMs, the future states can be predictable by calculating probability distributions over states at time $t+1$. Given probability distributions over states at $t$ and an action selected by the controller at $t$, probability distributions over state at $t+1$ or $t+k$ can be obtained by Eq.(\ref{eq:predict_state}) and (\ref{eq:predict_state_more}). As shown in Eq.(\ref{eq:predict_state_more_sub}), $\mathbf{P_t^*}$ is a marginal TPM, where $Prob(a_t=a(r))$ is set by uniform distribution such as $1/q\,\,\forall r$.
\begin{equation} \label{eq:predict_state}
        Prob_{pred}(S_{t+1})=\mathbf{P}_t^{a(r)}\cdot Prob(S_t)
\end{equation}
\begin{equation} \label{eq:predict_state_more}
        Prob_{pred}(S_{t+k})=\mathbf{P}_t^{*}\cdot Prob(S_{t+1})
\end{equation}
\begin{equation} \label{eq:predict_state_more_sub}
\begin{array}{ll}
        &\mathbf{P}_t^{*}=\{P_t^{*}(i.j)\}, 1\leq i,j \leq n_t, \textrm{where}\\
        &P_t^{*}(i,j)=\sum_{r=1}^{q}Prob\left(s_{t+1}=s(j)|s_t=s(i), a_t=a(r)\right)\\ 
        & \;\;\;\;\;\;\;\;\;\;\;\;\;\;\;\;\;\;\;\;\;\; \cdot \;Prob(a_t=a(r))
\end{array}
\end{equation}
\vspace{-0.5cm}
%#####################################################################################
\vspace{-0.2cm}
\subsection{Action-Reviser Module} \label{subsec:action-reviser}
\vspace{-0.2cm}
%#####################################################################################

This module inspects and revises an action $a_t$ chosen by a controller if required. The revision of chosen action is decided based on predicted probabilities of unfavorable states. Given $\mathcal{L}(S_t)$ and $a_t$, if predicted probabilities of the unfavorable states are higher than a threshold $\varrho_t$, $a_t$ will be revised. Otherwise, $a_t$ is applied.

The predicted probability distributions of states at $t+1$, $Prob_{pred}(S_{t+1})$, can be obtained by using Eq. (\ref{eq:predict_state}), where $\mathcal{L}(S_t)$, $Prob(S_t)$ and $P_t^{a(\cdot)}$ are provided from the e-FSM module, and the action $a(r)$ is chosen by a controller. The variant threshold $\varrho_t$ is set such that if a variance of $Prob_{pred}(S_{t+1})$ is larger, then $\varrho_t$ is smaller and vice versa as shown in Alg. \ref{alg:variantThreshold}.

\begin{algorithm}[h]
    \caption{$getVariantThreshold\,\,(Prob_{pred}(S_{t+1}))$}
    \label{alg:variantThreshold}
    \begin{algorithmic}[1]
    \Procedure{ $\varrho_t=f_{threshold}$}{$Prob_{pred}(S_{t+1})$}%\Comment{The g.c.d. of a and b}
        \State $X \gets Prob_{pred}(S_{t+1})$
        \State $X \gets$ $Sort$ $X$ by $descending$ $order$
        \State $E[X]=\sum_{j=1}^{n_t}j\cdot X(j)$
        \State $\varrho_t \gets X\left(floor\left(E[X]\right)\right)$
    \EndProcedure
\end{algorithmic}
\end{algorithm}

% two examples can be provided
For revising the chosen action, the action-reviser inspects other possible actions. For example, assuming that a given criterion for AV control is safety in a car-following scenario, an action $a_t$ is chosen by $1.0$, and the action-set (longitudinal acceleration) is defined by $\{-2.0,-1.0,0.0,1.0,2.0\} [m/s^2]$, the action-reviser inspects $a_t$ via calculating $Prob_{pred}(S_{t+1})$ and $\varrho_t$. If the predicted probability of the unfavorable states (e.g. collision states) are larger than $\varrho_t$, $a_t$ will be considered as an inappropriate action and replaced by one of other actions in the action-set which is slower than 1.0.

%#####################################################################################
\section{A Target Scenario and the DDPG Controller Settings}
\label{section:Target Scenario and Controller}
\vspace{-0.2cm}
%#####################################################################################
%#####################################################################################
The DDPG based controller is applied within the on-line evolving framework. In this section, a target car-following scenario and the DDPG based controller settings are discussed. The control performances w/ and w/o the on-line evolving framework under the scenario will be compared in Section \ref{section:experimental results}.

% #################################################################################
\vspace{-0.2cm}
\subsection{Target Car-Following Scenario}
\begin{figure}[ht]
\begin{center}
\includegraphics[width=7.5cm]{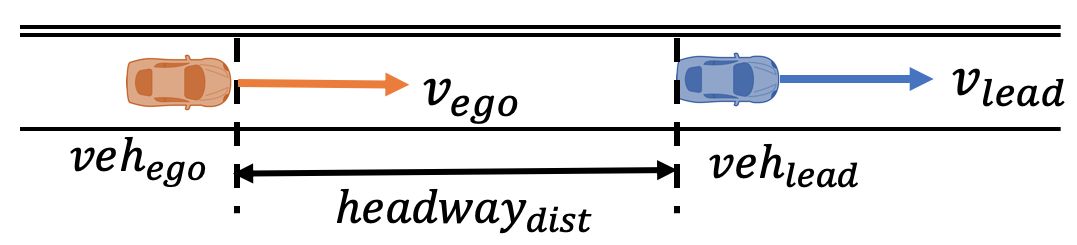}    % The printed column width is 8.4 cm.
\vspace{-0.3cm}
\caption{A target car-following scenario} 
\label{fig:car_following_scenario}
\end{center}
\end{figure}

\vspace{-0.2cm}
% #################################################################################
A target scenario is car-following on a single lane road as shown in Fig. \ref{fig:car_following_scenario}. The ego vehicle ($veh_{ego}$) and lead vehicle ($veh_{lead}$) are initialized by randomly assigning speeds and positions as far as the initial distance between two vehicles ($headway_{dist}$) is less than $100[m]$. The ranges of acceleration and velocity for the both vehicles are set by $[-2.0\,\,2.0] [m/s^2]$ and $[0.0\,\,\,32.0] [m/s]$ respectively. The ego vehicle's acceleration is chosen by the DDPG controller at every simulation time-step $t$=$0.25secs$, while the lead vehicle's acceleration is assigned by randomly selecting 200 secs (800 steps) of velocity profile from the 11 hours of realistic driving profile.
% explain about velocity profile (how does it obtain, how long..)
The vehicle model in the simulation is a point-mass model defined as Eq.(\ref{eq:point_mass_vehicle_model}) where $x_t(veh_i), \dot{x}_t(veh_i),$ and $\ddot{x}_t(veh_i)$ are position, velocity, and acceleration of $veh_i$ at time $t$.
\vspace{0.1cm}
\begin{equation} 
    \label{eq:point_mass_vehicle_model}
    \begin{array}{ll}
            &x_{t+1}(veh_i)=x_t(veh_i)+\dot{x_t}(veh_i)\cdot \Delta t \\
            &\dot{x}_{t+1}(veh_i)=\dot{x}_t(veh_i)+\ddot{x}_t(veh_i)\cdot \Delta t
    \end{array}
\end{equation}

The goal of the scenario is controlling the ego vehicle to follow the lead vehicle as $fast$ and $safe$ as possible. The episode is terminated in the following three cases: simulation step reaches the maximum step $800$, the $headway_{dist}$ is longer than $200[m]$, and the ego vehicle collides to the lead vehicle.

% #################################################################################
\vspace{-0.15cm}
\subsection{DDPG Controller}
\vspace{-0.2cm}
% #################################################################################
The DDPG is implemented as the ego vehicle's controller. It is one of the RL with MLP controllers, of which policy is represented by multilayered neural network. The policy is trained in a way to get the maximum value at every state given a pre-defined reward-function. As proposed by \cite{van2016deep}, the fixed size $\omega$ of histories are stored and implemented to train a policy of the RL with MLP controller; the stored histories can be represented by $\mathcal{H}_t=\{h_{t-\omega+1}, ..., h_t\}$ where $h_t=[\bar{z}_t, a_t, \bar{z}_{t+1}, \eta_t]$; $\bar{z}_t, \bar{z}_{t+1}, a_t,$ and $\eta_t$ are current and next states, an action selected by the controller, and a reward. Especially, the DDPG controller consists of the actor $\mu(\bar{z}|\theta^{\mu})$ and critic  $Q(\bar{z},a|\theta^Q)$ networks and implements ``soft-update" instead of directly copying weights of the networks in updating target-networks; see descriptions of \cite{lillicrap2015continuous} for details about the DDPG controller.

In a similar way in \cite{nageshrao2019interpretable}, the DDPG controller is configured for the given car-following scenario as the following: the ego vehicle's acceleration is set by $accel_t(veh_{ego})=\mu(\bar{z}_t|\theta^{\mu})+\mathcal{N}$, where $\bar{z}_t$ is a set of observations defined as Eq.(\ref{eq:state of ddpg1}), and $\mathcal{N}$ is the temporally correlated exploration noise generated by using the Ornstein-Uhlenbeck noise process proposed by \cite{uhlenbeck1930theory}. Also, the Adam method is applied for training the networks, as suggested by \cite{kingma2014adam}. Learning rates for the actor and critic networks are set by $10^{-4}$ and $10^{-3}$ respectively, and a discount factor is set by 0.95. 

% #################################################################################
\vspace{-0.2cm}
\begin{equation} \label{eq:state of ddpg1}
\begin{array}{ll}
        \bar{z}_t=[v_t(veh_{ego}), headway_{dist}, v_t(veh_{lead}), \\ \;\;\;\;\;\;\;\;\;accel_{t-1}(veh_{ego})]^T
\end{array}
\end{equation}
%#####################################################################################

Since the given control criteria in the scenario is $speed$ and $safety$, \cite{nageshrao2019interpretable} designs a reward function to minimize velocity difference and distance between two vehicles, and variation of ego vehicle's acceleration as shown in Eq.(\ref{eq:ddpg_reward_function1}-\ref{eq:ddpg_reward_function3}), where $headway_{const}$ and $d_{safe}$ are a constant headway-time and a minimum safe distance. The reward $\eta_t$ is defined by sum of $\eta_t^{vel}, \eta_t^{dist},$ and $\eta_t^{accel}$. 
%#####################################################################################
\begin{equation} \label{eq:ddpg_reward_function1}
       \eta_t^{vel}=e^{-\frac{(v_t(veh_{ego})-v_t(veh_{lead}))^2}{v_{max}}}-1
\end{equation}
\begin{equation}\label{eq:ddpg_reward_function2}
\eta_t^{dist}=e^{-\frac{(headway_{dist}-headway_{const}\cdot d_{safe})^2}{2\cdot headway_{const}\cdot d_{safe}}}-1\\
\end{equation}
\begin{equation}\label{eq:ddpg_reward_function3}
       \eta_t^{accel}=e^{-\frac{(\Delta accel(veh_{ego})^2}{2\cdot accel_{max}}}-1
\end{equation}

%#####################################################################################
\section{The On-line Evolving Framework Settings with DDPG controller}
\label{section:DDPG_controller_eFSM}
\begin{figure}[ht]
\begin{center}
\includegraphics[width=7.5cm]{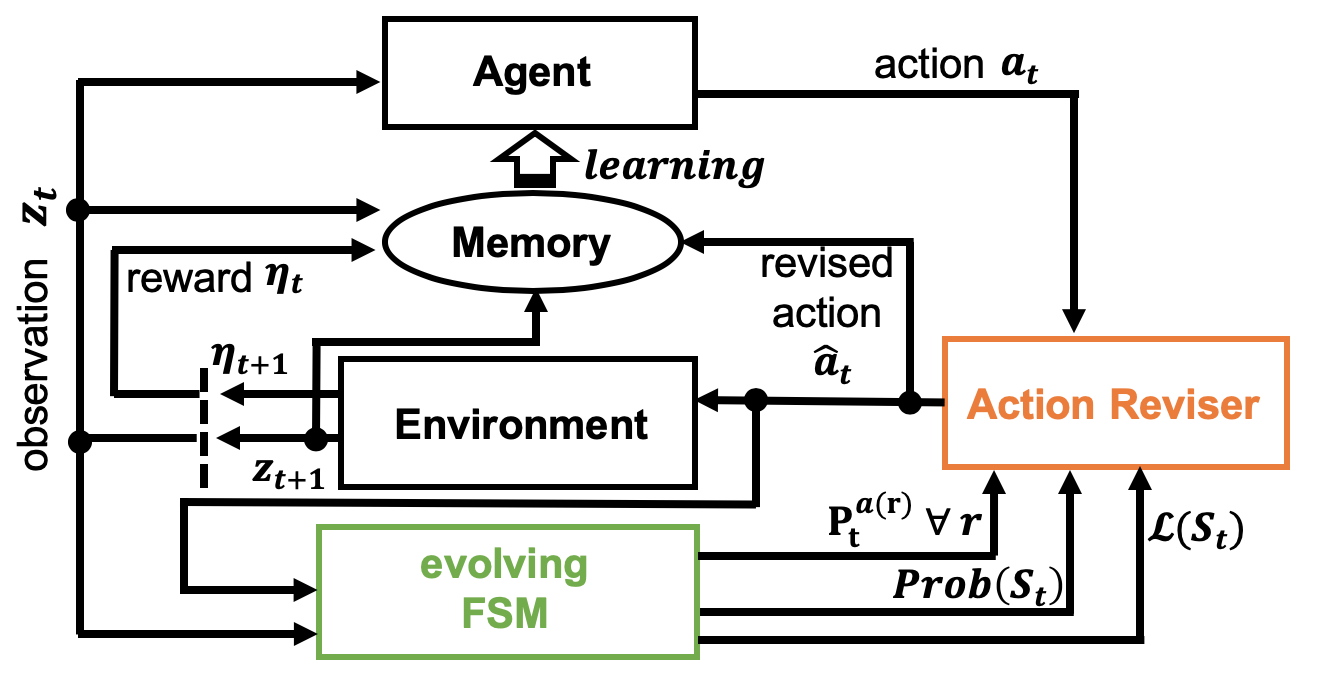}    % The printed column width is 8.4 cm.
\vspace{-0.2cm}
\caption{An on-line evolving framework with DDPG} 
\label{fig:RL_eFSM_framework}
\end{center}
\end{figure}

% \vspace{-0.2cm}
%#####################################################################################
The on-line evolving framework with the RL with MLP controller can be represented by Fig. \ref{fig:RL_eFSM_framework}. In this section, the configurations of e-FSM and action-reviser modules are discussed for the given car-following scenario.
%overview of online evolving framework with DDPG explain!
%In this section, we would like to explain how the e-FSM and action-reviser modules are configured to control the ego-vehicle under the car-following scenario.
%#####################################################################################
\vspace{-0.35cm}
\subsection{e-FSM Configurations}
\vspace{-0.2cm}
%#####################################################################################
As we discussed, the AV control criteria in the given scenario are $safety$ and $speed$. In simulating the scenario, the DDPG controller chooses longitudinal accelerations via its policy at every time $t$, and the e-FSM determines states and identifies state-transitions independently. Among the states determined by the e-FSM, some will be flagged as the unfavorable states, which represent either unsafe (collision) or slow-speed driving situation ($headwawy_{dist}$ is longer than $200[m]$). At every time-step, $s_t(i)$, where $\{s_t(i)\in S_t| i=argmax_i(Prob\left(s_t(i)\right)\}$, is flagged by:
\vspace{0.1cm}
\[\mathcal{L}(s_t(i)) = \left\{
  \begin{array}{ll}
    &l_n (none): neither\,\,crash\,\,nor\,\,large\textrm{-}distance\\
    &l_s\,\,(unfavorable_{safety}): crash\\
    &l_{ld} (unfavorable_{speed})\,\,\,: large\textrm{-}distance
  \end{array}
\right.
\]

As shown in Eq.(\ref{eq:state_eFSM}), three observations are selected to represent states in the e-FSM. The coefficients of eTS within the e-FSM module, $\rho$ and $\epsilon$, are set by 0.7 and 0.3 respectively. Since the DDPG controller uses the continuous action-space, the continuous action-set is encoded to the discrete action-set by using an interval $\delta=0.2 [m/s^2]$. The continuous action set $A_c=[-2.0\,\,2.0][m/s^2]$ is given for the both vehicles in the scenario, and encoded to the $q=20$ number of discrete actions.
\vspace{0.1cm}
\begin{equation} \label{eq:state_eFSM}
    z_t=\{v_t(veh_{ego}), headway_{dist}, v_t(veh_{lead})\}
\end{equation}
% such that $A_d=\{a(1), ..., a(20)\}$ where $a(1)=[-2.0\,\,-1.8)$, $...$, $a(19)=[1.6\,\,\,1.8)$, $a(20)=[1.8\,\,\,2.0]$. 

%#####################################################################################
\subsection{Action-Reviser Configurations}
\vspace{-0.2cm}
%#####################################################################################
If it is predicted that the selected action $a_t$ causes a transition to one of the unfavorable states, the action-reviser explores and selects another action as discussed in section \ref{subsec:action-reviser}. 

The DDPG controller chooses an action $a_t \in A_c$ at every time-step, and $a_t$ is encoded to a discrete action $a(r) \in A_d$. Then, future probability distributions over states are calculated via Eq.(\ref{eq:predict_state}) to inspect $a(r)$. In the predicted probability distributions, if any of the unfavorable states ($l_s$ or $l_{ld}$) have a higher probability than the dynamic threshold $\varrho_t$, the action-reviser revises $a(r)$. Otherwise, $a_t$ is returned and applied. The action-reviser continues exploring a discrete action space $(A_d)$ until an appropriate action is found. Specific steps are shown in Alg. \ref{alg:actionreviser}.

\begin{algorithm}[b]
    \caption{$actionsReviser$}
    \label{alg:actionreviser}
    \begin{algorithmic}[1]
    \Procedure{$\hat{a}_t=f_{AR}$}{$Prob(S_{t}), a_t, A_d, \textbf{P}_t^{a(\cdot)}, \mathcal{L}(S_t)$}%\Comment{The g.c.d. of a and b}
        \State{$r \gets encode(a_t, A_d)$}
        \State $Prob_{pred}(S_{t+1}) \gets \mathbf{P}_t^{a(r)}\cdot Prob(S_t)$
        \State {$I \gets f_{IA}\left(Prob_{pred}(S_{t+1}), \mathcal{L}(S_t)\right)$} \algorithmiccomment{Alg. 3}
        \If{$I=0$}  \algorithmiccomment{$l_s$ is expected}
            \While{$I=0 \And r>0$}
                \State{$r \gets r-1$}
                \State{$Prob_{pred}(S_{t+1}) \gets \mathbf{P}_t^{a(r)}\cdot Prob(S_t)$}
                \State{$I \gets f_{IA}\left(Prob_{pred}(S_{t+1}), \mathcal{L}(S_t)\right)$}
            \EndWhile
            \State{$\bar{a}_t \gets decode(a(r),A_d)+\hat{\mathcal{N}}(0,V)$}
            \State{\Return $\hat{a}_t \gets min(max(\bar{a}_t,a_{min}),a_{max})$}
        \ElsIf{$I=1$}   \algorithmiccomment{$l_{ld}$ is expected}
            \While{$I=1 \And r<q$}
                \State{$r \gets r+1$}
                \State{$Prob_{pred}(S_{t+1}) \gets \mathbf{P}_t^{a(r)}\cdot Prob(S_t)$}
                \State{$I \gets f_{IA}\left(Prob_{pred}(S_{t+1}), \mathcal{L}(S_t)\right)$}
            \EndWhile
            \State{$\bar{a}_t \gets decode(a(r),A_d)+\hat{\mathcal{N}}(0,V)$}
            \State{\Return $\hat{a}_t \gets min(max(\bar{a}_t,a_{min}),a_{max})$}
        \Else   \algorithmiccomment{$l_{n}$ is expected}
            \State{\Return $\hat{a}_t \gets a_t$}
        \EndIf
    \EndProcedure
\end{algorithmic}
\end{algorithm}

\begin{algorithm}
    \caption{$inspectAction$}
    \label{alg:inspectAct}
    \begin{algorithmic}[1]
    \Procedure{$I=f_{IA}$}{$Prob_{pred}(S_{t+1}), \mathcal{L}(S_t)$}%\Comment{The g.c.d. of a and b}
        \State $\varrho_t \gets f_{threshold}\left(Prob_{pred}(S_{t+1})\right)$ \algorithmiccomment{Alg. 1}
        \For{$i=1,2, ..., n_t$}
            \If{$Prob_{pred}(s_{t+1}(i))\geq\varrho$}
                \If{$\mathcal{L}(s_{t+1}(i))=l_s$}
                    \State{\Return $I\gets0$}
                \ElsIf{$\mathcal{L}(s_{t+1}(i))=l_{ld}$}
                    \State{\Return $I\gets1$}
                \Else
                    \State{\Return $I\gets2$}
                \EndIf
            \EndIf
        \EndFor
    \EndProcedure
\end{algorithmic}
\end{algorithm}

After action-revising process, $a(r)$ is decoded to a continuous action $\hat{a}_t$ by calculating a mean of corresponded range since each discrete action represents an individual partial range of the continuous action-set. For example, if the selected discrete action is $a(r)=[0.2\,\, 0.4)$, $0.3$ is returned as a revised action $\hat{a}_t$. In addition, the Gaussian noise with a decreasing variance $\hat{\mathcal{N}}(0,V)$ is defined as shown in Eq.(\ref{eq:revised_action}) and implemented to explore the action space for identifying the e-FSM's TPMs. 
\vspace{0.1cm}
\begin{equation} \label{eq:revised_action}
    \begin{array}{ll}
    \textrm{where}\,\,\,\,V=\frac{a_{max}}{max\left(1, (K*Iter_{episode})\right)}
    \end{array}
\end{equation}

Initially, $V$ is set by the maximum acceleration of the vehicle, but it is decreased over the number of the episode iterations. Note $K, a_{max}, a_{min},$ and $Iter_{episode}$ are an invariant constant, the maximum/minimum accelerations of the vehicle, and the number of episode iterations.
Since the e-FSM does not have any states in the beginning, appropriate inspection and revision of the chosen action are not possible. Thus, the action-reviser module is activated after the $\xi$ number of episode iterations; $K$ and $\xi$ are set by 0.001 and 50 in this study.

%#####################################################################################
\vspace{-0.1cm}
\section{Experimental Results}
\label{section:experimental results}
\vspace{-0.2cm}
%#####################################################################################
The car-following control performances of the DDPG w/ and w/o the on-line evolving framework are compared in terms of $safety$ and $speed$ criteria. In the beginning of every first episode, the DDPG and e-FSM modules are initialized such that weights of critic and actor networks in the DDPG are randomly assigned, the state-set of e-FSM is set by empty-set $S_0=\{\}$, and $F_o^{a(r)}(0)$ and $F^{a(r)}(0)$ are set by $\bar{\epsilon}$ for all $r$. While running the scenario repeatedly, the critic and actor networks are updated, and the e-FSM derives a stochastic model by determining states and identifying state-transitions over time.

\begin{figure}[h]
    \begin{center}
        \includegraphics[width=9.0cm]{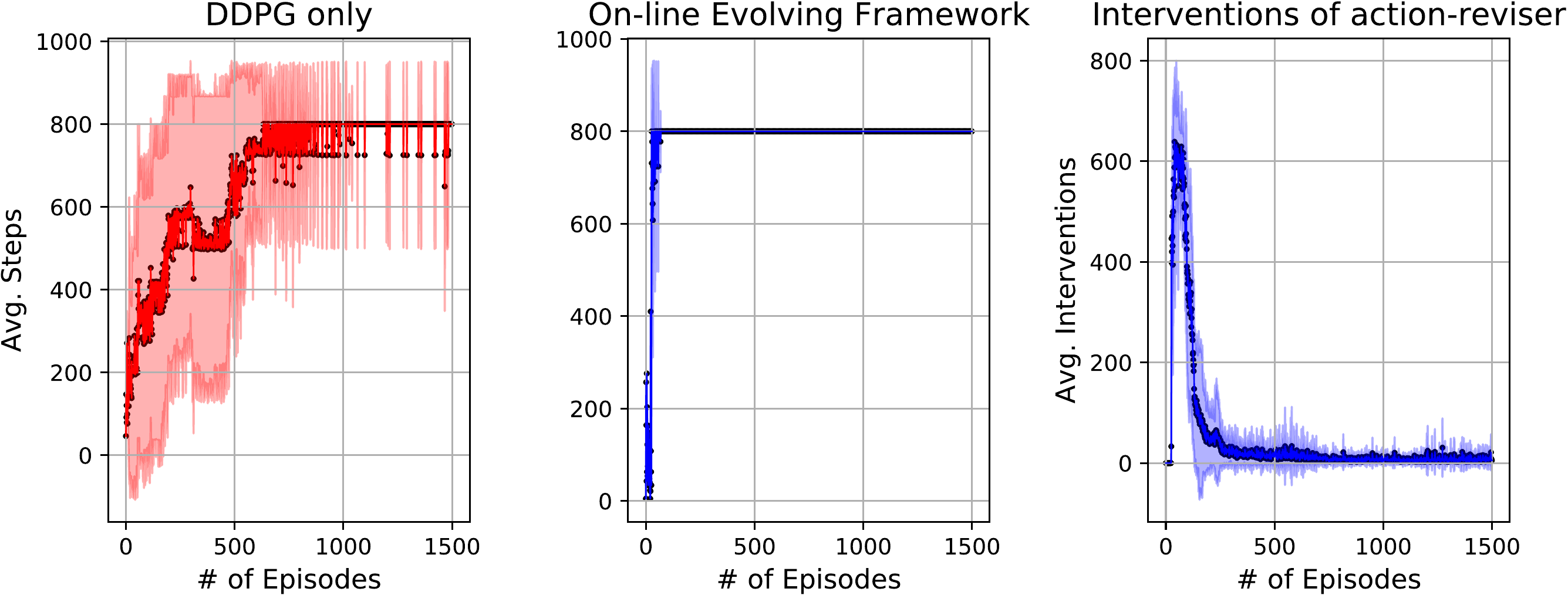}    % The printed column width is 8.4 cm.
        \vspace{-0.8cm}
        \caption{The control performance evolution in car-following by DDPG w/o (left) and w/ on-line evolving framework (center). The averaged number of interventions for revising actions chosen by DDPG (right).} 
    \label{fig:car_following_result_steps}
    \end{center}
\end{figure}

% over 10 different runs
Since every episode is terminated when the controller fails to avoid collisions or maintain $headway_{dist}$ less than $200[m]$, the controller's performance can be analyzed by comparing the total number of simulated steps in each episode, where the maximum step is 800. The Fig. \ref{fig:car_following_result_steps} plots the evolution of the car-following performances for DDPG w/ and w/o the on-line evolving framework averaged over 10 different runs; 1500 episodes are iterated in each run. The different number of succeeded and failed episodes (in total 15000 episodes) are observed depending on the implementation of the proposed framework as below table.  
\begin{table}[h]
    \centering
    \begin{tabular}{||c | c | c | c||}
        \hline
        Type & Success & Large-distance & Collision \\ [0.5ex] 
        \hline\hline
        DDPG & 12249 (81.6\%) & 1251 & 1500 \\ 
        \hline
        Our Approach & 14719 (98.1\%) & 76 & 205 \\
        \hline
    \end{tabular}
\end{table}
\vspace{-0.3cm}

As shown in Fig. \ref{fig:car_following_result_steps}, the control performances of the both approaches are enhanced over iterations. But, the DDPG w/ the on-line evolving framework (proposed approach) achieves the given control-goal (driving safe and fast) with fewer iterations. Also, our approach does not fail after 68 iterations, but the DDPG w/o the framework does. it is observed the number of interventions for revising actions is decreasing as DDPG policy is more optimized; there are no interventions during the first 50 episodes because the action-reviser is inactivated. These results imply that the DDPG controller's wrong decision-making is detected and revised appropriately within our proposed framework. Besides, it is noticed that a stochastic model is properly created by the e-FSM module, enabling precise prediction of future states based on the choice of actions.
% Also, it can be possible by predicting future states precisely via a stochastic model which is created by the e-FSM module.

% explain how much faster and longer steps can be obtained.
% considering providing stochastic data for speed difference in safe cases.
% consider why DDPG only case is slow or not gauranteed its performance 
% why evolving framework case is not perfect : one step prediction is limited.
Since the available maximum speed of ego vehicle depends on the lead vehicle's speed in the car-following, the best way to satisfy the speed criterion is driving as fast as the lead vehicle. Thus, the speed difference between ego and lead vehicles among succeeded episodes is quantified. The overall mean and variance of the velocity difference are 2.0460 and 1.7800 for the DDPG with the on-line evolving framework, while 1.9497 and 1.8057 for the DDPG only. 

Although the DDPG w/ the on-line evolving framework controls the ego vehicle slightly slower than the DDPG only, it achieves both criteria, safety and speed, more successfully by trading off priority of the two criteria. 

%#####################################################################################
\vspace{-0.2cm}
\section{Discussion and Conclusion}
\label{section:conclusion}
\vspace{-0.2cm}
%#####################################################################################
In this work, the on-line evolving framework is proposed to advance the decision-making capability of controllers. Since the e-FSM module can create a stochastic model by evolving the model's structure and identifying state-transition dynamics through experiences, it becomes possible to predict future states based on the choice of actions. The action-reviser module checks if an action chosen by the controller causes unfavorable situations in the future. And, if one of the unfavorable situations is highly expected, it explores the given action-space to choose another action. 

In the given car-following scenario, DDPG w/ and w/o the on-line evolving framework are implemented to control the ego vehicle as $fast$ and $safe$ as possible. As shown in the experimental results, the crashes and large-distances does not occur with our proposed framework after few episode iterations whereas the control failures are continuously occurred w/o the framework. It shows that the e-FSM generates a stochastic model precisely which enables to detect incorrect decisions effectively and revise them appropriately within the on-line evolving framework. 
\vspace{-0.3cm}
\bibliography{ifacconf}             % bib file to produce the bibliography

\begin{thebibliography}{16}
\providecommand{\natexlab}[1]{#1}
\providecommand{\url}[1]{\texttt{#1}}
\providecommand{\urlprefix}{URL }
\expandafter\ifx\csname urlstyle\endcsname\relax
  \providecommand{\doi}[1]{doi:\discretionary{}{}{}#1}\else
  \providecommand{\doi}{doi:\discretionary{}{}{}\begingroup
  \urlstyle{rm}\Url}\fi

\bibitem[{Al-Qizwini et~al.(2017)Al-Qizwini, Barjasteh, Al-Qassab, and
  Radha}]{al2017deep}
Al-Qizwini, M., Barjasteh, I., Al-Qassab, H., and Radha, H. (2017).
\newblock Deep learning algorithm for autonomous driving using googlenet.
\newblock In \emph{2017 IEEE Intelligent Vehicles Symposium (IV)}, 89--96.
  IEEE.

\bibitem[{Bojarski et~al.(2016)Bojarski, Del~Testa, Dworakowski, Firner, Flepp,
  Goyal, Jackel, Monfort, Muller, Zhang et~al.}]{bojarski2016end}
Bojarski, M., Del~Testa, D., Dworakowski, D., Firner, B., Flepp, B., Goyal, P.,
  Jackel, L.D., Monfort, M., Muller, U., Zhang, J., et~al. (2016).
\newblock End to end learning for self-driving cars.
\newblock \emph{arXiv preprint arXiv:1604.07316}.

\bibitem[{Filev and Kolmanovsky(2010)}]{filev2010markov}
Filev, D.P. and Kolmanovsky, I. (2010).
\newblock Markov chain modeling approaches for on board applications.
\newblock In \emph{Proceedings of the 2010 American Control Conference},
  4139--4145. IEEE.

\bibitem[{Gadepally et~al.(2013)Gadepally, Krishnamurthy, and
  Ozguner}]{gadepally2013framework}
Gadepally, V., Krishnamurthy, A., and Ozguner, U. (2013).
\newblock A framework for estimating driver decisions near intersections.
\newblock \emph{IEEE Transactions on Intelligent Transportation Systems},
  15(2), 637--646.

\bibitem[{Han et~al.(2019{\natexlab{a}})Han, Filev, and
  Ozguner}]{han2019online}
Han, T., Filev, D., and Ozguner, U. (2019{\natexlab{a}}).
\newblock An online evolving framework for modeling the safe autonomous vehicle
  control system via online recognition of latent risks.
\newblock \emph{arXiv preprint arXiv:1908.10823}.

\bibitem[{Han et~al.(2019{\natexlab{b}})Han, Jing, and
  {\"O}zg{\"u}ner}]{han2019driving}
Han, T., Jing, J., and {\"O}zg{\"u}ner, {\"U}. (2019{\natexlab{b}}).
\newblock Driving intention recognition and lane change prediction on the
  highway.
\newblock In \emph{2019 IEEE Intelligent Vehicles Symposium (IV)}, 957--962.
  IEEE.

\bibitem[{Kingma and Ba(2014)}]{kingma2014adam}
Kingma, D.P. and Ba, J. (2014).
\newblock Adam: A method for stochastic optimization.
\newblock \emph{arXiv preprint arXiv:1412.6980}.

\bibitem[{Kuge et~al.(2000)Kuge, Yamamura, Shimoyama, and Liu}]{kuge2000driver}
Kuge, N., Yamamura, T., Shimoyama, O., and Liu, A. (2000).
\newblock A driver behavior recognition method based on a driver model
  framework.
\newblock Technical report, SAE Technical Paper.

\bibitem[{Kurt and {\"O}zg{\"u}ner(2008)}]{kurt2008hybrid}
Kurt, A. and {\"O}zg{\"u}ner, {\"U}. (2008).
\newblock Hybrid state system development for autonomous vehicle control in
  urban scenarios.
\newblock \emph{IFAC Proceedings Volumes}, 41(2), 9540--9545.

\bibitem[{Lillicrap et~al.(2015)Lillicrap, Hunt, Pritzel, Heess, Erez, Tassa,
  Silver, and Wierstra}]{lillicrap2015continuous}
Lillicrap, T.P., Hunt, J.J., Pritzel, A., Heess, N., Erez, T., Tassa, Y.,
  Silver, D., and Wierstra, D. (2015).
\newblock Continuous control with deep reinforcement learning.
\newblock \emph{arXiv preprint arXiv:1509.02971}.

\bibitem[{Mnih et~al.(2015)Mnih, Kavukcuoglu, Silver, Rusu, Veness, Bellemare,
  Graves, Riedmiller, Fidjeland, Ostrovski et~al.}]{mnih2015human}
Mnih, V., Kavukcuoglu, K., Silver, D., Rusu, A.A., Veness, J., Bellemare, M.G.,
  Graves, A., Riedmiller, M., Fidjeland, A.K., Ostrovski, G., et~al. (2015).
\newblock Human-level control through deep reinforcement learning.
\newblock \emph{Nature}, 518(7540), 529.

\bibitem[{Nageshrao et~al.(2019{\natexlab{a}})Nageshrao, Costa, and
  Filev}]{nageshrao2019interpretable}
Nageshrao, S., Costa, B., and Filev, D. (2019{\natexlab{a}}).
\newblock Interpretable approximation of a deep reinforcement learning agent as
  a set of if-then rules.
\newblock In \emph{2019 18th IEEE International Conference On Machine Learning
  And Applications (ICMLA)}, 216--221. IEEE.

\bibitem[{Nageshrao et~al.(2019{\natexlab{b}})Nageshrao, Tseng, and
  Filev}]{nageshrao2019autonomous}
Nageshrao, S., Tseng, H.E., and Filev, D. (2019{\natexlab{b}}).
\newblock Autonomous highway driving using deep reinforcement learning.
\newblock In \emph{2019 IEEE International Conference on Systems, Man and
  Cybernetics (SMC)}, 2326--2331. IEEE.

\bibitem[{Redmill et~al.(2008)Redmill, Ozguner, Biddlestone, Hsieh, and
  Martin}]{redmill2008ohio}
Redmill, K.A., Ozguner, U., Biddlestone, S., Hsieh, A., and Martin, J. (2008).
\newblock Ohio state university experiences at the darpa challenges.
\newblock \emph{SAE International Journal of Commercial Vehicles},
  1(2008-01-2718), 527--533.

\bibitem[{Uhlenbeck and Ornstein(1930)}]{uhlenbeck1930theory}
Uhlenbeck, G.E. and Ornstein, L.S. (1930).
\newblock On the theory of the brownian motion.
\newblock \emph{Physical review}, 36(5), 823.

\bibitem[{Van~Hasselt et~al.(2016)Van~Hasselt, Guez, and Silver}]{van2016deep}
Van~Hasselt, H., Guez, A., and Silver, D. (2016).
\newblock Deep reinforcement learning with double q-learning.
\newblock In \emph{Thirtieth AAAI conference on artificial intelligence}.

\end{thebibliography}
                                                   % with bibtex (preferred)
                                                   
%\begin{thebibliography}{xx}  % you can also add the bibliography by hand

%\bibitem[Able(1956)]{Abl:56}
%B.C. Able.
%\newblock Nucleic acid content of microscope.
%\newblock \emph{Nature}, 135:\penalty0 7--9, 1956.

%\bibitem[Able et~al.(1954)Able, Tagg, and Rush]{AbTaRu:54}
%B.C. Able, R.A. Tagg, and M.~Rush.
%\newblock Enzyme-catalyzed cellular transanimations.
%\newblock In A.F. Round, editor, \emph{Advances in Enzymology}, volume~2, pages
%  125--247. Academic Press, New York, 3rd edition, 1954.

%\bibitem[Keohane(1958)]{Keo:58}
%\newblock \emph{Power and Interdependence: World Politics in Transitions}.
%\newblock Little, Brown \& Co., Boston, 1958.

%\bibitem[Powers(1985)]{Pow:85}
%\newblock Is there a way out?
%\newblock \emph{Harpers}, pages 35--47, June 1985.

%\bibitem[Soukhanov(1992)]{Heritage:92}
%\newblock \emph{{The American Heritage. Dictionary of the American Language}}.
%\newblock Houghton Mifflin Company, 1992.

%\end{thebibliography}

% \appendix
% \section{A summary of Latin grammar}    % Each appendix must have a short title.
% \section{Some Latin vocabulary}              % Sections and subsections are supported  
                                                                         % in the appendices.
\end{document}